\documentclass[aps,pra,  groupedaddress]{revtex4}
\usepackage{graphicx,subfigure}
\usepackage{epsfig}

\usepackage{bbold}
\usepackage{amsmath}
\usepackage{amsthm}
\usepackage{amsfonts}
\usepackage{amssymb}
\usepackage{enumerate}
\newcommand{\E}{\mathbb{E}}

\begin{document}
\title{Separability Criterion for Multi-Mode Gaussian States }
\author{K. V. S. Shiv Chaitanya }
\email[]{ chaitanya@hyderabad.bits-pilani.ac.in}
\affiliation{Department of Physics, BITS Pilani, Hyderabad Campus, Jawahar Nagar, Shamirpet Mandal,
Hyderabad, India 500 078.}
\author{Sibasish Ghosh}
\email[]{ sibasish@imsc.res.in}
\affiliation{Optics $\&$ Quantum Information Group,The Institute of Mathematical Sciences, C.I.T Campus, Taramani, Chennai, India, 600113.}

\author{V Srinivasan}
\email[]{ vsspster@gmail.com}
\affiliation{Department of Theoretical Physics, University of Madras,  Guindy, Chennai, India, 600025.}

\begin{abstract}
In this paper, we give Separability criterion for  the multi-mode Gaussian states using the Marchenko-Pastur theorem. We show that the Marchenko-Pastur theorem from random matrix theory as necessary and sufficient condition for separability of multimode Gaussian states.
\end{abstract}

\maketitle
\section{introduction}
In the theory of quantum information, the separability problem, namely, to find out weather an arbitrary state of a given  bipartite system is entangled or separable is a quantum state is a NP hard \cite{np}. There are several techniques to detect the entanglement of a given quantum state and one of them is the positive partial transposition (PPT) criterion. 
This was first proposed by Peres  for the finite dimensions  as a necessary condition \cite{peres},
and later Horodecki's  showed this condition to be necessary and
sufficient for separability in the case of  $2 \times 2$ and $2 \times 3$ dimensions
 \cite{horodecki}. In literature, this condition is known as Peres-Horodecki criterion.  In the infinite dimensional cases, for the class of Gaussian states, the  Peres-Horodecki criterion two mode case has been given  by Simon \cite{simon} using the consept of wigner distribution.  As the dimension of the system increases  testing  separability becomes difficult using PPT technique as there are PPT entangled states in such cases.  To overcome this drawback, one of the ways is to study this problem by using the
techniques of random matrix theory.  

In finite dimensional case, the PPT criterion for random quantum states is a well studied subject in the literature. This problem was first been studied numerically in the ref \cite{ns} and
the  random matrix and free probability techniques are applied to the study the linear maps \cite{hu}. Wishart distribution for the random states used in the ref \cite{hu1}. For more details of application of random matrices to the quantum information is given in \cite{hu1}. In this paper, we would like to address the issue of separability of the multi-mode Gaussian states using the Marchenko-Pastur theorem  \cite{mar}.

The Gaussian states play a very important role in quantum optics and quantum information. The best example of the Gaussian states are the squeezed states which are produced in the laboratory. It is well known that multimode squeezed state are entangled. As the modes of the Gaussian states increases it becomes difficult to test the entanglement criterion. The entanglement criterion other than  bipartite is tripartite exist  \cite{ado1} which is an example of multimode case. In general for N-mode case it is still an open problem but by treating the system as bipartite system  with N-m modes and m modes than for this case entanglement criterion of N symmetric  Gaussian states exist \cite{ado}. In this paper, we give entanglement criterion for the ensemble of n-mode system  with with N-m modes and m modes as bipartite system using the Merchnko Pasture law in random matrix theory. The Marchenko-Pastur law appears in the famous Marchenko-Pastur theorem \cite{mar}. 

In the case of the Gaussian states their exist an equivalent to representation in terms of density operator in a Hilbert space in terms of the Wigner function in a phase space.  The Wigner function  captures all the features of the wave function.This is a  quasi-probability distributions are well used in optics, to distinguish between the classical and non-classical properties of light. The non-classical proprieties of light are exhibited by the squeezed states. Hence, the information content of the given quantum system can be studied in terms of the Wigner function and it can become an important testing tool for the separability. The Wigner function is completely defined terms of the covariance matrix and condition for multi-mode squeezing is derived in \cite{gau,gau1,gau2,simon}. In literature, G S Agarwal has used the random matrices to  analyze the effects of dissipation in quantum optics \cite{aga}.

Even in the case of the Gaussian states as the number of modes increases, the dimension of the covariance matrix increases this gives rise to the same challenges as in the case of finite dimension. To get over this problem we will use the random matrix techniques. In random matrix theory, the Wishart matrices are characterized by  covariance matrix and the probability distribution of the ensemble is given by the Laguerre  polynomials.  Then the empirical distribution of eigenvalues of the covarience matrix satisfy Marchenko-Pastur theorem \cite{mar}. In this paper, we show that it is necessary and sufficient that the eigenvalues of the multi-mode covariance matrix lies within the limits of Marchenko-Pastur law in order that state is separable, and if atleast one of the eigenvalues is negative then it will fall outside the limit of the Marchenko-Pastur law. Or, in other words, the Marchenko-Pastur law gives the convergence of the eigenvalues of the covariance matrix.

\section{Random Matrix Theory}
In random matrix theory, the dynamics of the ensemble of an infinite dimensional random matrix $H$ is described by the probability distribution function \cite{ml}
\begin{equation}\label{pdfo}
{\cal P}(\lambda_1,\ldots,\lambda_N)\,d\Lambda
= c_ne^{-\beta H}
\,\,d\Lambda,
\end{equation}
here, by diagonalizing  the random matrix $H$ with a suitable orthogonal, unitary, or symplectic matrix   is brought into the following form $H=-\sum_{i=1}^NV(\lambda_i)-
\beta\sum_{i<1}^Nln\vert\lambda_i-\lambda_j\vert$ where $V(\lambda_i)$ is the potential
and the $\lambda_i$ are the eigenvalues with $i$ being a free index running from $1,2..N$,  then $\vert\lambda_i-\lambda_j\vert$ is the Vandermonde determinant,   $d\Lambda=d\lambda_1 \ldots d\lambda_N$, $c_n$ is constant of proportionality and the index $\beta=1, 2, 4$ characterises the real parameters of the symmetry class of orthogonal, unitary and symplectic respectively for the random matrix $H$.

 As the random matrix is invariant under the symmetry group these ensembles are called the invariant ensembles. In the random matrix $H$,  $V(\lambda_i)$ is the potential, $\lambda_i$ are the eigenvalues with $i$ being a free index running from $1,2,\cdots N$ and $\vert\lambda_i-\lambda_j\vert$ is the Vandermonde determinant. For details please refer to \cite{ml,rm, ran1,ran2,ran3, ran4}. The probability distribution function can be rewritten as 
\begin{equation}\label{pdf}
{\cal P}(\lambda_1,\ldots,\lambda_N)\,d\Lambda
=c_n e^{-\sum_{i=1}^N\beta V(\lambda_i)}
\prod_{i<j}\vert\lambda_i-\lambda_j\vert^\beta
\,\,d\Lambda
\end{equation}
By manipulating the Vandermonde determinant, that is, by adding and deleting columns or rows, the equation (\ref{pdf}) is written as
\begin{equation}\label{pdf1}
{\cal P}(\lambda_1,\ldots,\lambda_N)\,d\Lambda
=c_n \prod_{i=1}^N w^{\frac{1}{2}} (\lambda_i)
\prod_{i<j}\vert\lambda_i-\lambda_j\vert
\,\,d\Lambda
\end{equation}
where $w_\beta (\lambda_i)$ is the weight function of classical orthogonal polynomials.
The classical orthogonal polynomials are classified into three different 
categories depending upon the range of the polynomials. The polynomials in the intervals $(-\infty;\infty)$ with weight function $w=e^{-\lambda^2}$ are the Hermite polynomials.
In the intervals $[0;\infty)$ with weight function $w=\lambda^be^{-\lambda}$ are the Laguerre polynomials. In the intervals $[-1;1]$ with weight function $w=(1+\lambda)^{a}(1-\lambda)^b$ are the Jacobi polynomials.

Hence, the invariant ensembles in random matrix theory is classified into three class, the Gaussian ensembles these  matrices are known as the Wigner matrices whose probability distribution function of the ensemble  is given by the Hermite polynomials,  the Wishart matrices whose probability distribution function of the ensemble is given by the Laguerre  polynomials and the two Wishart matrices whose probability distribution function of the ensemble is given by the Jacobi  polynomials. For details ref \cite{rm}. The Marcenko Pastur's Quarter-Circle Law is given for the  Wishart matrices and they are defined as follows.

Namely, let ($X_1,\cdots, X_n$) be $m\times n$ matrix with  independent and identically distributed random column vectors $X_i$ in $\mathcal{R}^m$ with Gaussian distribution. Here, $m$ and $n$ corresponds to  the number of variables $m$ and the number of observations $n$ or $m$ is the sample size and $n$ is the dimension of the vectors. The Gaussian distribution for $X_i$ are characterized by  mean $0$ and covariance matrix $I_m$. Then one can construct an empirical covariance matrix $\Sigma$ with the dimension $m\times m$, as follows 
\begin{eqnarray}\label{grm}
\Sigma=\frac{1}{n}\sum_{i=1}^n(X_i-\bar{X})(X_i-\bar{X})^T_i=\frac{1}{n}YY^T.
\end{eqnarray}
The  probability distribution function of the random covariance matrix $\Sigma$ is computed in terms of its eigenvalues and 
is given in terms of  the Laguerre polynomials \cite{rm}. As one has the following propriety 
 $E(\frac{1}{n}YY^⊤)=I_m$, then, from the strong law of large numbers as $n$ tends to infinity it indicates that with probability one,
\begin{eqnarray}
lim_{n\rightarrow\infty}\frac{1}{n}YY^T=I_m.
\end{eqnarray}

In the case  $n$ grows with $m$, (here $n$ is the number of random variables and the $m$ is the dimension of the Hilbert space) one has the following theorem known as the Marchenko-Pastur theorem \cite{mar}. 

Theorem: \textit{Let us assume for simplicity that the components of $X_i$ are i.i.d. Gaussian random variables with zero mean, unit variance, and bounded moments that is there is some bound B, independent 
 of m, such that $\forall n, E(\vert x_{i j}\vert^k) \leq B$.
Then n depends on m in such a way that $m/n \rightarrow r \leq 1$ as
$n\rightarrow\infty$.
Under these assumptions, the distribution of the eigenvalues of $\frac{1}{n}YY^T$ asymptotically
approaches the Marcenko-Pastur law as $n\rightarrow\infty$
\begin{equation}
f(x) =\frac{\sqrt{(x-a)(b-x)}}{2\pi xr}.
\end{equation}}
where $a = (1-\sqrt{r})^2$
and $b = (1+\sqrt{r})^2$. 

For the case $r=1$ it reduces to a famous quarter circle law.
In this paper, we use this theorem to test separability Criterion for multi-Mode Gaussian States. The best example for Gaussian states are the squeezed states. 
 
In proving  Marcenko-Pastur law it is assumed that the entries are i.i.d. but this case corresponds to the zero discord, therefore we would like to relax the condition of independence. It has been shown in the paper \cite{gy} for the case where a class of random matrices with dependent entries whose empirical distribution of eigenvalues satisfies  Marcenko-Pastur law. For completeness we state the theorem again here as defined in the paper \cite{gy}

Definition[Condition {\bf C0}] \label{def:C0}
Let $\{A_n\}_{n \geq 1}$ be a sequence of $n \times N$ real random matrices where $N = N(n)$ and $c_n := N/n$.  We let $r_1^{(n)}, \ldots, r_n^{(n)}$ denote the rows of $A_n = (\zeta_{ij}^{(n)})_{1 \leq i \leq n, 1 \leq j \leq N}$ and define the $\sigma$-algebra associated to row $k$ as
$$ \mathcal{F}_k^{(n)} := \sigma( r_1^{(n)}, \ldots, r_{k-1}^{(n)}, r_{k+1}^{(n)}, \ldots, r_n^{(n)} ) $$
for all $k = 1, \ldots, n$.  Let $\E_k[ \cdot ]$ denote the conditional expectation with respect to the $\sigma$-algebra associated to row $k$.  We then say that the sequence $\{A_n\}_{n \geq 1}$ obeys condition {\bf C0} if the following hold:
\begin{enumerate}[(i)]
\item $\E_k[ \zeta_{ki}^{(n)} ] = 0$ for all $i,k,n$ \label{C0:uncor}
\item One has
$$ q_n := \sup_{k} \frac{1}{n} \sum_{i=1}^N \E|\E_k[ (\zeta_{ki}^{(n)})^2 ] - 1| = o(1) $$
 \label{C0:var}
\item One has 
$$ \sup_{k, i \neq j} |\E_k[ \zeta_{ki}^{(n)} \zeta_{kj}^{(n)}]| + \sup_{k,i,j \neq l} |\E_k[ (\zeta_{ki}^{(n)})^2 \zeta_{kj}^{(n)} \zeta_{kl}^{(n)}]| = O(n^{-1/2} \gamma_n) $$
a.s., where $\gamma_n \rightarrow 0$ as $n \rightarrow \infty$. \label{C0:2cor}
\item $\sup | \E_k[ \zeta_{ki}^{(n)} \zeta_{kj}^{(n)} \zeta_{kl}^{(n)} \zeta_{km}^{(n)}]| = O(n^{-1} \gamma_n)$ a.s where the supremum is over all $k$ and all $i,j,l,m$ distinct. \label{C0:4cor}
\item $\sup_{n,i,j} \E |\zeta_{ij}^{(n)}|^4 \leq M < \infty$ \label{C0:moment}
\item One has
$$ \rho_n := \sup_{k} \frac{1}{n^2} \sum_{1 \leq i,j \leq N} \E|\E_k[ (\zeta_{ki}^{(n)})^2 (\zeta_{kj}^{(n)})^2] - 1| = o(1). $$ \label{C0:row_variance}
\item There exists a non-negative integer sequence $\beta_n = o(\sqrt{n})$ such that $\sigma(r_{i_1}^{(n)}, \ldots, r_{i_k}^{(n)})$ and $\sigma(r_{j_1}^{(n)}, \ldots, r_{j_m}^{(n)})$ are independent $\sigma$-algebras whenever
$$ \min_{1 \leq l \leq k, 1 \leq p \leq m} |i_l - j_p| > \beta_n. $$ \label{C0:indep}
\end{enumerate}

Condition \eqref{C0:uncor} implies that the entries from different rows of the given random matrix are uncorrelated while \eqref{C0:2cor} and \eqref{C0:4cor} allow for a weak correlation amongst entries in the same row.  Condition \eqref{C0:var} is a requirement on the variance of the entries and \eqref{C0:moment} is a moment assumption on the entries.  Condition \eqref{C0:row_variance} is of a technical nature.  In particular, \eqref{C0:row_variance} (along with \eqref{C0:var}) allows one to control terms of the form
$$ \sup_{k} var \left( \frac{1}{n} |r_k^{(n)}|^2 \right) $$
where $|r_k^{(n)}|$ is the Euclidian norm of the vector $r_k^{(n)}$.  In words, condition \eqref{C0:indep} implies that rows, which are ``far enough apart,'' are independent.  

This case corresponds to multi-mode a system if one only considered the nearest neighbor interaction the best example is spin chains. The techniques of random matrix  is applied to spin chain is well studied subject.

\section{Gaussian States}

 Consider a n-mode Gaussian state in the density matrix representation
\begin{equation}\label{sta}
\rho=\rho_1\otimes\rho_2\otimes\cdots\otimes\rho_n
\end{equation}
where $\rho_i$ are the two mode Gaussian states.
Then the Wigner function of $n$ mode Gaussian state in phase space is given by
\begin{equation}\label{wig}
W(\rho)=\frac{exp[-\frac{1}{2}(X-\bar{X})^T V^{-1}(X-\bar{X})]}{(2\pi)^n \sqrt{det[V]}}.
\end{equation}
where $X$ is a vector defined in $2m$ dimensional phase space variables 
\begin{eqnarray*}
X = 
\left( \begin{array}{cccccccccccc}
	q_{A1}&
	\cdots &
	q_{An}&
	q_{B1}&
	\cdots &
	q_{Bn}&
	p_{A1}&
	\cdots &
	p_{An}&
	p_{B1}&
	\cdots &
	p_{Bn}
	   \end{array}
	\right),
\label{eq:eq3}
\end{eqnarray*} 
or in terms of hermitian canonical operators
\begin{eqnarray*}
\hat{X} = 
\left( \begin{array}{cccccccccccc}
	\hat{q}_{A1}&
	\cdots &
	\hat{q}_{An}&
	\hat{q}_{B1}&
	\cdots &
	\hat{q}_{Bn}&
	\hat{p}_{A1}&
	\cdots &
	\hat{p}_{An}&
	\hat{p}_{B1}&
	\cdots &
	\hat{p}_{Bn}
	   \end{array}
	\right),
\label{eq:eq31}
\end{eqnarray*} and $V$ is called the covarience matrix, will be defined subsequently.   
These phase space variables $X$ or $\hat{X} $ satisfy the following commutation relationship \cite{gau2}
\begin{eqnarray}
[\hat{X}_{\alpha}, \hat{X}_{\beta}] & = & i\,\Omega_{\alpha \beta},
~~~ \alpha, \beta = 1, 2, 3, 4; \nonumber \\
\Omega & = & \left( \begin{array}{cccc}
			J & \cdots &\cdot & 0\\
			\cdot &\cdot &\cdot &\cdot \\
		    \cdot &\cdot &\cdot &\cdot \\
			0 & \cdots &\cdot &J
		    \end{array} \right),\;\;\; J = \left( \begin{array}{cc}
						0 & 1\\
						-1 & 0
						\end{array}\right),
\label{eq:eq4}
\end{eqnarray}
here $\Omega$ is a $4n\times 4n$ symplectic matrix. 
A pure state is a Gaussian if and only if its Wigner function is non-negative. The Gaussianity of a N-mode 
wigner function is defined in terms of its first and second moments. The first moments are defined as
\begin{equation}
\langle \hat{X}\rangle=Tr(\hat{X}\rho)
\end{equation}
which is called mean value which can be made zero.
The second moment is known as the covarience matrix $V$, is defined as
\begin{equation}\label{cob}
V_{ij}=\frac{1}{2}\langle \{\Delta\hat{x}_i,\Delta\hat{x}_j\}\rangle, 
\end{equation}
where $\Delta 
\hat{X} = \hat{X} - \langle \hat{X} \rangle$. The knowledge of the covariance matrix alone is enough to describe the N-mode Gaussian state, hence
the entangled properties of the system doesn't dependent on the mean value $\langle \hat{X}\rangle$ and purely determined in terms of covariance matrix $V$.

Given an any Wigner function, the  condition for squeezing is characterized in terms of covariance matrix and it should satisfy uncertainty relationship \cite{gau,gau1, gau2}
\begin{equation}
V + \frac{i}{2}\,\Omega  \geq 0,
\label{unce}
\end{equation}
which implies $V > 0$, for details ref \cite{simon, gau1, gau2}.
For the two mode Gaussian state, the associated covariance matrix is $4\times 4$, separability criterion  is equivalent to the 
Peres-Horodecki separability criterion \cite{simon}. This also holds for the $(1+N)$ mode Gaussian states, with one mode is under the possession of Allice and rest $N$ modes are under possession of Bob \cite{addo}.
 
 In this paper we are interested in the separability of  genuine multi-partite $2n$- mode  Gaussian states with $2n$ of far apart parties, each having in his /her possession a single mode. On the other hand for the bipartite scenario, there will be only two far apart parties, say Alice and Bob, each of which is in possession of $n$ modes each, or we can have something intermediate.

\section{Result}
To derive the condition of separability of multi-mode Gaussian states, we consider a state as defined in equation (\ref{sta}), and  we apply partial transposition criterion. The state considered here is $n$-mode Gaussian state, in this state we consider only the nearest neighbour interaction. For applying partial transposition criterion we divide this $n$-modes to $m$-modes and $(n-m)$-modes, here $m$ is strictly less than $n$.  The density matrix $\rho$ given in the equation (\ref{sta}) is in the infinite dimensional Hilbert space, from this space we go over to the
 the phase space and the state is described by the Wigner function (\ref{wig}),  this Wigner function is totally characterized by the covariance matrix (\ref{cob}), defined in the previous section. To bring the covariance matrix to the canonical form we symplectic transformation $ S^{(r)}$ acting on $2n$ modes which preserves the commutation relations \cite{gau2} and it is element of group $SP(2n, R)$.
The real linear transformation on $\hat{X}$ is given in terms of 
$4n\times 4n$ real matrix, namely $ S^{(r)}$
\begin{equation}
\hat{X}\rightarrow\hat{X}'=S^{(r)}\hat{X}
\end{equation}
 We apply $S^{(r)}$ in such a way that the $\hat{X}'$ is given as
\begin{eqnarray*}
\hat{X}' = 
\left( \begin{array}{cccccccccccc}
	\hat{q}_{A1}&
	\hat{q}_{B1}&
	\cdots &
	\hat{q}_{An}&
	\hat{q}_{Bn}&
	\hat{p}_{A1}&
	\hat{p}_{B1}&
	\cdots &
	\hat{p}_{An}&
	\hat{p}_{Bn}&
	   \end{array}
	\right).
\label{eq:eq32}
\end{eqnarray*}
In terms of annihilation and creation operators which is equivalent to
\begin{eqnarray*}
\hat{X}' = 
\left( \begin{array}{cccccccccccc}
	\hat{a}_{A1}&
	\hat{b}_{B1}&
	\cdots &
	\hat{a}_{An}&
	\hat{b}_{Bn}&
	\hat{a}^\dagger_{A1}&
	\hat{b}^\dagger_{B1}&
	\cdots &
	\hat{a}^\dagger_{An}&
	\hat{b}^\dagger_{Bn}&
	   \end{array}
	\right).
\label{eq:eq23}
\end{eqnarray*}
Thus the covariance matrix is written as 
\begin{eqnarray}\label{cov}
V=\left( \begin{array}{cccc}
			\sigma_{11} & \cdots &\cdot & \sigma_{1n}\\
			\cdot &\cdot &\cdot &\cdot \\
		    \cdot &\cdot &\cdot &\cdot \\
			\sigma_{n1} & \cdots &\cdot &\sigma_{nn}
		    \end{array} \right),
\end{eqnarray}
where $\sigma_{ij}$ is a $2\times 2$ matrix constructed from expectation values of  $(a_{Ai},b_{Bi},a^\dagger_{Aj},b^\dagger_{Bj},)$ elements. Since $V$  is a bonifide  
the covariance matrix of Gaussian states given by equation (\ref{sta}),  the eigenvalue distribution of this covariance matrix should follow the
Marchnko Pasture law.  

By identifying this covariance matrix Gaussian states with the Gaussian random covariance matrix given in equation (\ref{grm}), we get that $m=2n$ and $n=4n$. One can clearly see that as $m\rightarrow\infty$ and $n\rightarrow\infty$
the ration $m/n=1/2$. Therefore, the distribution of the eigenvalues of $V=\hat{\xi}'\hat{\xi}'^T$ asymptotically
approaches the Marcenko-Pastur law as $m,n\rightarrow\infty$
\begin{equation}
f(x) =\frac{\sqrt{(x-a)(b-x)}}{\pi x}.
\end{equation}
where $a = (1-\sqrt{\frac{1}{2}})^2$
and $b = (1+\sqrt{\frac{1}{2}})^2$.

To verify the separability, we divide the $n$-modes into two groups 
$m$- modes and $(n-m)$-modes, here $m<n$, and apply partial transposition on the $(n-m)$-modes.  After partial transposition the new covariance matrix is given by $\tilde{V}$, the new covariance matrix is bonifide covariance matrix for the given Gaussian state than it has to  satisfies the following uncertainty relationship

\begin{equation}
\tilde{V}+\frac{i}{2}\,\Omega \geq 0, \;\;\; \tilde{V} = \Lambda V \Lambda,
\end{equation}
where $\Lambda=(1,\cdots,1,1,-1,\cdots,1,-1)$, there are total $4n$ ones,
the first $2n$ the $+1$ corresponds to for the $q_{Ai},q_{Bi}$ in the second $2n$ the $+1$ corresponds to the $p_{Ai}$ and$p_{Bi}$, as the partial transposition is applied in the  $k+1\cdots n$, all the quadratures of  $p_{Bi}$
from $k+1$ mode to  $n$ mode will flip the sign to  $-1$ . In the covariance matrix this is equivalent to transposing each $\sigma_{ij}$.
Hence one has 
\begin{eqnarray}\label{cov1}
\tilde{V}=\left( \begin{array}{cccccc}
			\sigma_{11} & \cdots &\sigma_{1k}& \sigma_{1k+1}^T&\cdot & \sigma_{1n}^T\\
			\cdot &\cdot &\cdot &\cdot &\cdot &\cdot  \\
		    \cdot &\cdot &\cdot &\cdot &\cdot &\cdot \\
		    \cdot &\cdot &\cdot &\cdot &\cdot &\cdot  \\
		    \cdot &\cdot &\cdot &\cdot & \cdot &\cdot \\
			\sigma_{n1}^T & \cdots &\sigma_{nk+1}^T& \sigma_{1k}&\cdot &\sigma_{nn}
		    \end{array} \right),
\end{eqnarray}
Then the Peres-Horodecki criterion for the multimode continuous variables or the Simon's criterion for multimode reduces to 
the following theorem. 

Theorem : \textit{It is  necessary and sufficient condition for separability for a given  multimode Gaussian state if the eigenvalue distribution of the partially transposed covariance matrix $\tilde{V}$  satisfies Marcenko-Pastur law .}

Proof: Let us assume that the partial transposed covariance matrix $\tilde{V}$ is a  bonifide  
covariance matrix of Gaussian states and the eigenvalue distribution of this covariance matrix satisfy's the
Merchnko Pasture law.  Thus by identifying this covariance matrix of Gaussian states with the Gaussian random covariance matrix we get that $m=2n$ and $n=4n$. One can clearly see that as $m\rightarrow\infty$ and $n\rightarrow\infty$ the ratio $m/n=1/2$. Therefore, the distribution of the eigenvalues of $V=\hat{\xi}'\hat{\xi}'^T$ asymptotically approaches the Marcenko-Pastur law as $m,n\rightarrow\infty$
\begin{equation}
f(x) =\frac{\sqrt{(x-a)(b-x)}}{\pi x}.
\end{equation}
where $a = (1-\sqrt{\frac{1}{2}})^2$
and $b = (1+\sqrt{\frac{1}{2}})^2$. 
This gives a condition that the eigenvalues  $x$ is bounded above and below by $2\sqrt{2}<x-3<-2\sqrt{2}$. Marcenko-Pastur law is also known as semi-circle law in random matrix theory, for m = n,
that is distribution of all the eigenvalues of the random matrix fall on a semi-circle, In our case m 6= n, hence it
will not follow a semi-circle law but the eigenvalues are bounded. It is also proved in random matrix theory that
the eigenvalues of the covariance matrix are positive \cite{rm}, if state is entangled, it is known from the Peres-Horodecki
criterion or equivalently Simon criterion Gaussian states that atleast one of the eigenvalue of the covariance matrix
has to be negative. Hence, if the state is entangled then at least one of the eigenvalue is negative it will be outside the
semi-circle that is in the other semi-circle which will be in the negative direction. If all the eigenvalues are positive
then they will be within the range and the states are separable.

\section{Conclusion}
In this paper,  we give entanglement criterion for  the multi-mode Gaussian states using the Marchenko-Pastur theorem. We have shown that the Marchenko-Pastur theorem from random matrix theory as necessary and sufficient condition for separability.

\section*{Acknowledgments}
Authors thank KVSSC acknowledges the Department of Science and technology, Govt of India (fast track scheme (D. O. No: SR/FTP/PS-139/2012)) for financial support.


\end{document}